\documentclass[12pt]{iopart}

\usepackage[amssymb,cdot]{SIunits}
\usepackage{graphicx}
\usepackage{bm}
\usepackage{amssymb}
\usepackage{mathrsfs}
\usepackage{rotating}
\usepackage{supertabular}

\begin{document}

\title[XRR in the presence of surface contamination]{Limitations of X-ray reflectometry in the presence of surface contamination}


\author{D.L. Gil\footnote{Present address:
Department of Mechanical and Aerospace Engineering, Princeton University, Princeton, NJ} and D. Windover}
\address{National Institute of Standards and Technology, 100 Bureau Dr., Stop 8520, Gaithersburg, MD 20899-8520}
\ead{windover@nist.gov}

\begin{abstract}
Intentionally deposited thin films exposed to atmosphere often develop unintentionally deposited few monolayer films of surface contamination. This contamination arises from the diverse population of volatile organics and inorganics in the atmosphere.  Such surface contamination can affect the uncertainties in determination of thickness, roughness and density of thin film structures by X-Ray Reflectometry (XRR).  Here we study the effect of a \unit{0.5}{\nano\meter} carbon surface contamination layer on thickness determination for a \unit{20}{\nano\meter} titanium nitride thin film on silicon.  Uncertainties calculated using Markov-Chain Monte Carlo Bayesian statistical methods from simulated data of clean and contaminated TiN thin films are compared at varying degrees of data quality to study (1) whether synchrotron sources cope better with contamination than laboratory sources and (2) whether cleaning off the surface of thin films prior to XRR measurement is necessary.  We show that, surprisingly, contributions to uncertainty from surface contamination can dominate uncertainty estimates, leading to minimal advantages in using synchrotron- over laboratory-intensity data.  Further, even prior knowledge of the exact nature of the surface contamination does not significantly reduce the contamination's contribution to the uncertainty in the TiN layer thickness. We conclude, then, that effective and standardized cleaning protocols are necessary to achieve high levels of accuracy in XRR measurement.
\end{abstract}

\pacs{06.20.Dk, 61.05.cm, 68.55.jd, 68.35.Ct}
\submitto{JPD}

\maketitle


\section{Introduction}

X-ray reflectometry is widely used for characterizing the thickness, roughness, and density of nanometer scale thin films. Because it uses wavelengths of a similar or smaller scale relative to the thicknesses of the layers being studied, the resulting data has a relatively direct connection to the structure and therefore has a rather straightforward traceability to the International System of Units (SI) \cite{bowen98,bowen01,hasche03}. This is a significant advantage over other techniques like spectroscopic ellipsometry (SE), whose results are somewhat more difficult to interpret. But x-ray reflectometry is still somewhat sensitive to surface contamination. A comprehensive study by Seah et al. on ultra-thin SiO$_2$ layers on Si has shown offsets between x-ray reflectometry (XRR), neutron reflectometry (NR), spectroscopic ellipsometry, and x-ray photoelectron spectroscopy (XPS) \cite{seah_ultrathin_2003}. These offsets are believed likely due to the different effects contamination has for each technique. To use x-ray reflectometry for high-accuracy measurements of film thickness and roughness -- as the National Institute of Standards and Technology (NIST) plans to do for thin-film standards -- these effects must be quantified.

The primary difficulty in quantifying the effects of contamination is that thin surface contamination layers are hard to measure. The surface contamination with which we are concerned typically takes the form of rough, near-monolayer-thicknes, carbonaceous compounds. This is challenging to measure with x-ray or neutron reflectivity-based methods: carbon's low scattering factor for both techniques and the poor quality of the carbon layers produce low-contrast fringes.

X-ray photoelectron spectroscopy (XPS) and other inelastic scattering techniques can be used to measure the quantity per surface area -- and thus relative thicknesses -- of contamination layers, but generally require calibration in order to measure absolute thicknesses \cite{seah_ultrathin_2003}. This calibration is often, for denser materials, conducted by reflectometry measurements. In addition, these experiments are typically conducted in ultra-high-vacuum, which may change the properties, or even the very presence, of the adsorbed volatile contamination layers. This makes using these experiments somewhat challenging for achieving high accuracy in absolute thickness determination.

But a crucial question is whether these experiments are needed at all: How sensitive are other parameters determined by modeling of x-ray reflectometry data to the presence of an unknown, hard-to-measure layer of surface contamination? (E.g., how much do you have to know about contamination to achieve a given level of uncertainty in thickness for other layers within a structure?) And what sort of data quality (i.e., dynamic and $q$-space/angular range) is needed to achieve desired levels of uncertainties? (E.g., do you need synchrotron data?)

We study these questions by a simulation-based study of the effects of a carbon contamination layer on a TiN film on Si substrate structure being considered by NIST for an X-ray reflectometry standard. Using a Bayesian statistical approach to XRR data analysis, we estimate uncertainties for structural parameters of the high-Z TiN layer under various contamination and data quality conditions.

\section{Background}

\subsection{X-ray reflectometry}

X-ray reflectometry can be used to measure the density, thickness, and roughness of thin films which are laterally homogeneous at the scale of the beam \cite{chason97,pietsch04}. We limit ourselves here to the case of layered structures with fairly sharp interfaces and no density grading other than that provided by roughness. Density is measured by the critical angle for total external reflection and through the careful analysis of oscillation amplitudes; thickness is measured by the period of intensity oscillations appearing after the critical angle; and roughness is measured by the rapidity of overall intensity and oscillation intensity fall-off. The analysis here uses the Parratt recursion \cite{parratt54} with the perturbation for Gaussian roughness described by \cite{nevot80}. 

Analyzing XRR data is an inverse problem: because only the intensity -- rather than the complex amplitude -- of the reflected beam can be measured, in general there is no unique structure determined by an XRR pattern. The typical approach is to fit the parameters of a multilayer structure using an optimization approach, the most common being Genetic Algorithms (GAs) \cite{dane98, ulyanenkov05, wormington99}. Though optimization finds a best-fit solution -- and thus a best estimate of the parameter values -- it does not provide estimates of uncertainties on the parameters. To calculate uncertainties, a more sophisticated -- and computationally expensive -- approach is necessary.  NIST has developed statistical Markov Chain Monte Carlo (MCMC) methods in order to obtain parameter estimations and uncertainties within a Bayesian formalism \cite{windover_nist_2007}.

\subsection{Data analysis}

To make inferences about physical structure from XRR data by Genetic Algorithm (GA) or Markov Chain Monte Carlo (MCMC) methods requires a physical model that relates structural parameters to idealized non-noisy XRR data. This is provided by the Parratt recursion with Nevot-Croce roughness described above.

A physical model is not sufficient, however, because the data collected are (at best) noisy and (at worst) corrupted by systematic instrument effects. There must be, in addition, a statistical model of the data. For the GA method, this is a $\chi^2$ cost function; for the MCMC method, a probability density function. The cost function and probability density function employed here make very similar assumptions about the statistical characteristics of the data, see \cite{sivia96}. Nonetheless, the GA and MCMC methods each recover different types of information from the data.

X-ray reflectometry data consists of pairs of angles and measured intensities. For this study, we assume data free of angular errors with counting-error-limited measured intensities. The error in measured intensities is modeled using a log-normal likelihood with standard deviation of the square root of the calculated intensities; this approximates well a Poisson (i.e., counting-statistic) likelihood \cite{sivia96}.

We use a tiered data analysis architecture.  XRR model parameters are first determined for a given structural model using a GA optimization approach \cite{wormington99}.  In this analysis, we used a 1000 genome population evolved over 1000 generations to obtain best-fit structural parameters.  This structural information was then used to initialize the starting parameters for a tuned Markov Chain Monte Carlo (MCMC)  sampler to provide us with probability distributions for each parameter within a given structural model.  Each MCMC was allowed a 50 000 steps conditioning run to tune the MCMC target dimensions.  The MCMCs were then run for 250 000 steps to obtain adequate statistics for inter-comparison.  The tiered analysis was performed several times on each data set to validate the refinement stability.

\subsection{Simulated data}

\begin{figure}
  \resizebox{.45\textwidth}{!}
  	{\includegraphics{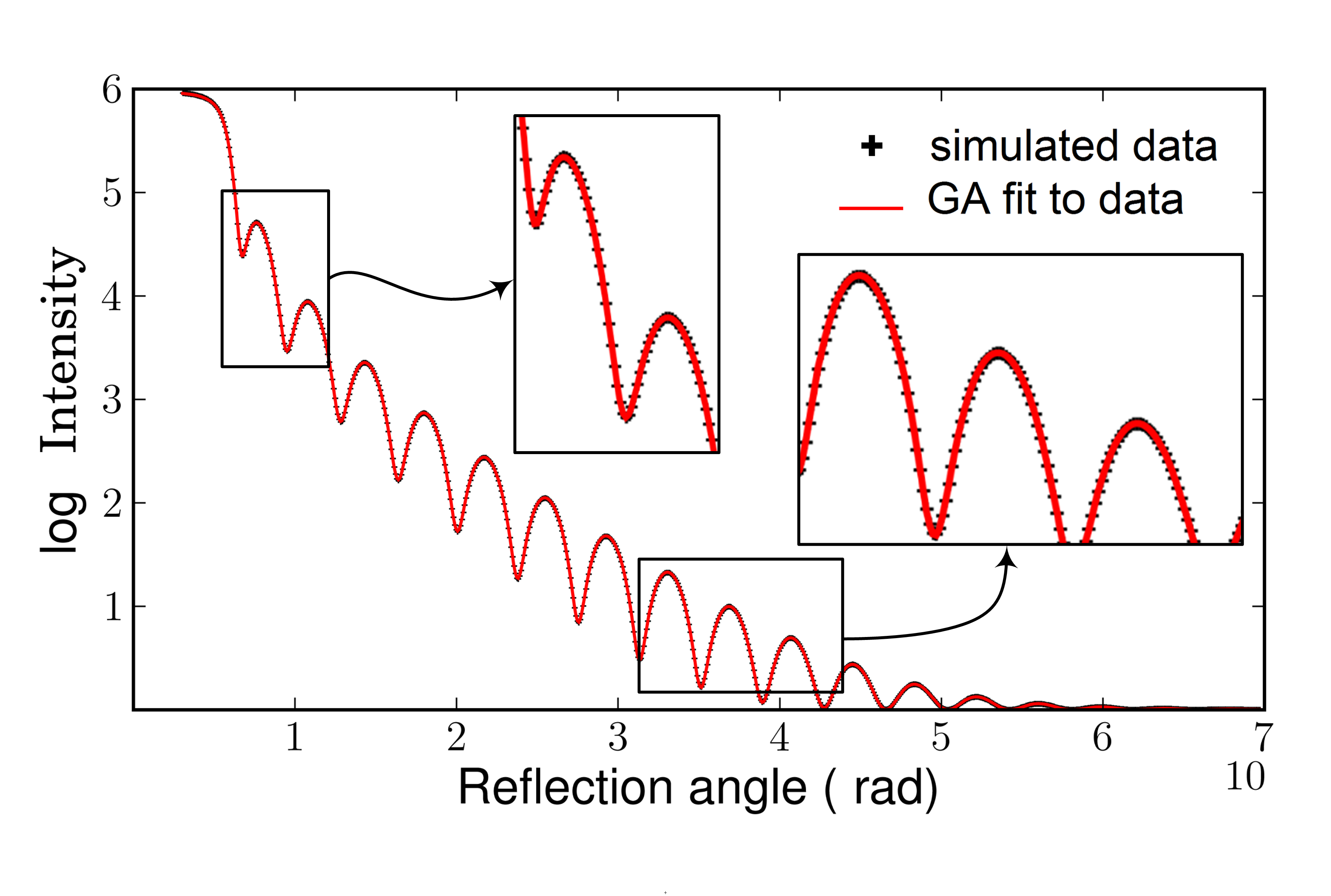}}
  \caption{Simulated (Cu radiation) X-ray reflectometry (XRR) data for case 2 structural modal and laboratory quality data (see Tables \ref{table:sim2} and \ref{table:quality}).  XRR simulated data (plus signs) has been fit using a genetic algorithm refinement (solid line) and yielded nearly identical parameters to those used in the simulation (Table \ref{table:sim2}).  Note refinement quality (magnified regions).}
\label{fig:XRR_lab}
\end{figure}

\begin{figure}
  \resizebox{.45\textwidth}{!}
  	{\includegraphics{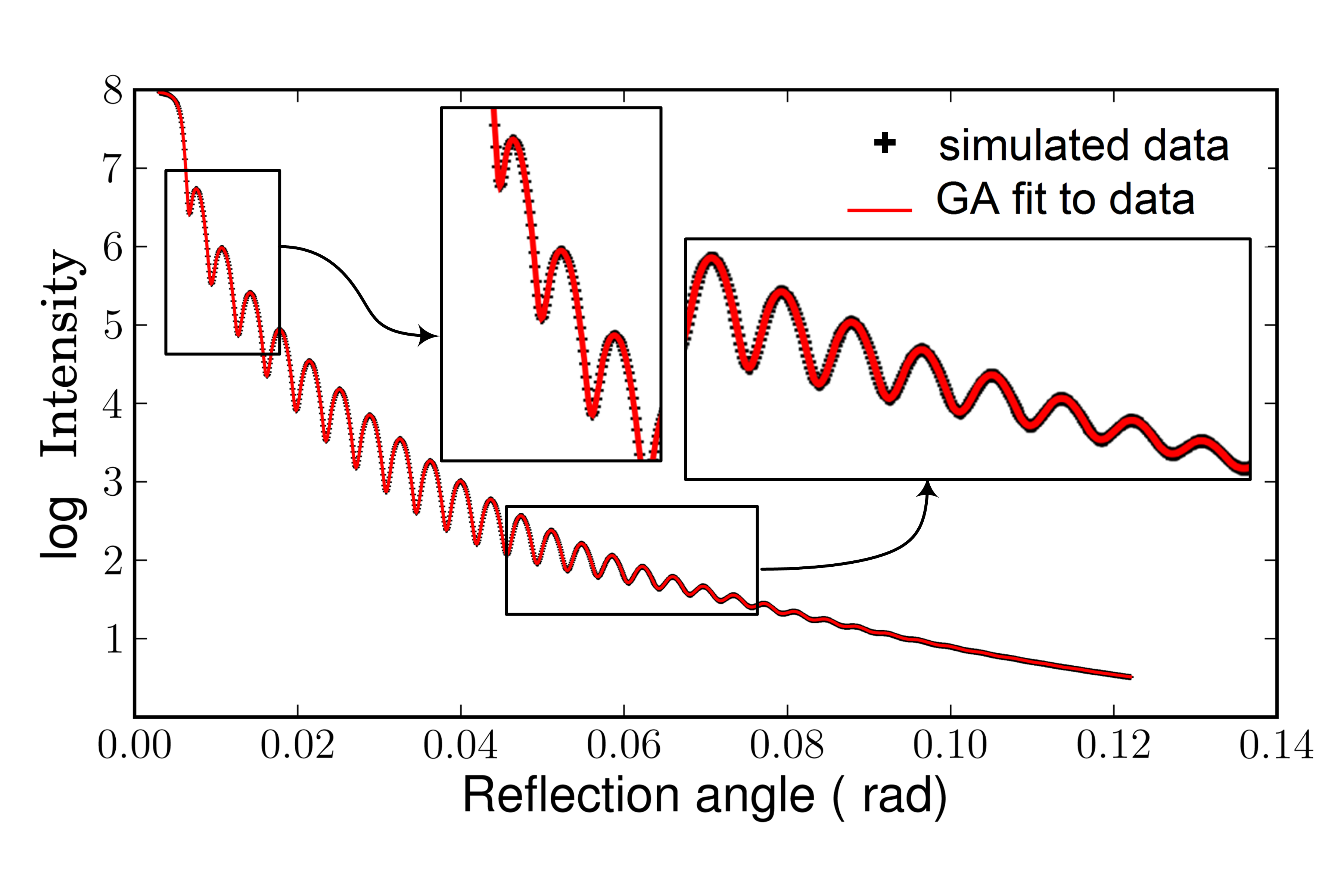}}
  \caption{Simulated (Cu radiation) X-ray reflectometry (XRR) data for case 2 structural modal and synchrotron quality data (see Tables \ref{table:sim2} and \ref{table:quality}).  XRR simulated data (plus signs) has been fit using a genetic algorithm refinement (solid line) and yielded nearly identical parameters to those used in the simulation (Table \ref{table:sim2}). Note refinement quality (magnified regions).}
\label{fig:XRR_syn}
\end{figure}

Two structures were simulated for this study:  case 1 -- a single layer of TiN on an Si substrate (for film parameters, see Table \ref{table:sim1}) -- and case 2 -- a contaminated single layer of TiN on an Si substrate (see Table \ref{table:sim2}).  Simulations were performed under two different data quality conditions (see Table \ref{table:quality}) selected to compare parameter refinement results between an advanced laboratory instrument and a synchrotron measurements from, e.g., a third-generation bending magnet beamline.  By way of illustration, we present XRR data simulations from our case 2 structure for laboratory (see Figure \ref{fig:XRR_lab} and synchrotron (see Figure \ref{fig:XRR_syn}) data quality conditions.  By considering these two cases, we can answer an oft-asked question for XRR data-collection: How many orders of magnitude data quality are needed to determine a given XRR model parameter?  The Bayesian statistical approach used here can directly answer this question by determining the `best-case' theoretically possible from XRR measurements for any refined parameter.


\begin{table}
\begin{tabular}{lcccc}
\hline
 & $t / \textrm{nm}$
 & $\sigma / \textrm{nm}$
 & $\rho / \textrm{g cm}^{-2}$
 \\
 \hline

 TiN & 20.0 & 0.5 & 4.90 \\
 Si & -- & 0.4 & 2.49 \\
 \hline
\end{tabular}
\caption{Case 1: clean TiN/Si structure. Composition, thickness ($t$), roughness ($\sigma$), and density ($\rho$).}
\label{table:sim1}
\end{table}

\begin{table}
\begin{tabular}{lcccc}
 \hline
 & $t / \textrm{nm}$
 & $\sigma / \textrm{nm}$
 & $\rho / \textrm{g cm}^{-2}$
 \\
 \hline
 C & 0.5 & 0.1 & 2.25 \\
 TiN & 20.0 & 0.5 & 4.90\\
 Si & -- & 0.4 & 2.49 \\
 \hline
\end{tabular}
\caption{Case 2: carbon-contaminated TiN/Si structure.}
\label{table:sim2}
\end{table}

\begin{table}
\begin{tabular}{lccccc}
 \hline
 & $2\theta$ & Step & Maximum & Background \\
 & range	& size & intensity & \\
 & (degrees) & (degrees) & (counts) & (counts)\\

 \hline
Laboratory case & 4.0 & 0.005 & $10^6$ & 1\\
Synchrotron case & 7.0 & 0.005 & $10^8$ & 1\\ 
\hline
\end{tabular}
\caption{Parameters of XRR data quality cases used in simulations.}
\label{table:quality}
\end{table}

\subsection{MCMC analysis}
The MCMC analysis method has initial optimal parameters input using the results of a GA.  Details of MCMC methods are beyond the scope of this paper.  The most important point is that all MCMC implementations, if properly tuned and allowed sufficient time, should produce the same result.  Most research into, and the complications in, MCMC methods relate to improving sampler efficiency and thus the number of samples required. Thus the details of the particular sampling scheme relate mainly to efficiency; the resulting samples are from the same Bayesian posterior probability distribution.

It is important for interpreting the probability distributions sampled by MCMC to know three modeling assumptions:  (1) the allowed prior ranges for each parameter within a model, (2) the assumed prior distributions for each parameter, and (3) the type of noise assumed within the data.  In this work, we use ranges for a uniform prior which assume the same physical structure (same number of layers) as the simulated data; the ranges of the priors are generous to allow the MCMC to sample a wide parameter space.  In Table \ref{table:model1} we give the allowed parameter ranges for case 1.  In Table \ref{table:model2}, we provide the ranges used in case 2.  In Table \ref{table:model2a}, we use a highly constrained model for case 2 with all the values of the carbon contamination layer fixed at their simulated values; i.e., we assume we know the nature of the contamination exactly.  In each model, we assume a uniform prior distribution for thickness, roughness, and density.  In all cases we assume the noise in actual measured data -- and thus the likelihood of the data for a given set of parameter values -- is Poisson.

\begin{table}
\begin{tabular}{lcccc}
 \hline
 & $t / \textrm{nm}$
 & $\sigma / \textrm{nm}$
 & $\rho / \textrm{g cm}^{-2}$
 \\
 \hline
TiN & 15.0 to 25.0 & 0.01 to 2.5 & 4.0 to 6.0 \\
Si & -- & 0.01 to 2.5 & 2.0 to 3.0 \\
\hline
\end{tabular}
\caption{Model 1: Allowed MCMC (uniform prior) ranges for TiN/Si structure with no surface contamination layer.}
\label{table:model1}
\end{table}

\begin{table}
\begin{tabular}{lcccc}
 \hline
 & $t / \textrm{nm}$
 & $\sigma / \textrm{nm}$
 & $\rho / \textrm{g cm}^{-2}$
 \\
 \hline
 C & 0.0 to 2.0 & 0.01 to 2.5 & 2.0 to 3.0 \\
 TiN & 15.0 to 25.0 & 0.01 to 2.5 & 4.0 to 6.0 \\
 Si & -- & 0.01 to 2.5 & 2.0 to 3.0 \\
 \hline
\end{tabular}
\caption{Model 2: Allowed MCMC (uniform prior) ranges for TiN/Si multilayer with surface contamination layer.}
\label{table:model2}
\end{table}

\begin{table}
\begin{tabular}{lcccc}
 \hline
 & $t / \textrm{nm}$
 & $\sigma / \textrm{nm}$
 & $\rho / \textrm{g cm}^{-2}$
 \\
 \hline
 C & 0.5 & 0.1 & 2.25 \\
 TiN & 15.0 to 25.0 & 0.01 to 2.5 & 4.0 to 6.0 \\
 Si & -- & 0.01 to 2.5 & 2.0 to 3.0 \\
 \hline
\end{tabular}
\caption{Model 2a: Allowed MCMC range for TiN/Si multilayer with a known surface contamination layer.}
\label{table:model2a}
\end{table}

\section{Results}

Statistically-determined uncertainties for laboratory and synchrotron levels of data quality have been calculated using the MCMC method for a simulated clean TiN sample (case 1) with corresponding modeling ranges (model 1)  and a simulated carbon contaminated TiN sample (case 2) with its corresponding modeling ranges (model 2).  Absolute and relative uncertainties for each structural parameter and each data quality are presented.  We also present a modified analysis for case 2, in which we provide the exact parameters for the contamination layer as prior information for the MCMC method (model 2a) and discuss the resulting uncertainties.

\subsection{Clean sample  -- case 1}

\begin{figure}
  \resizebox{.45\textwidth}{!}
  	{\includegraphics{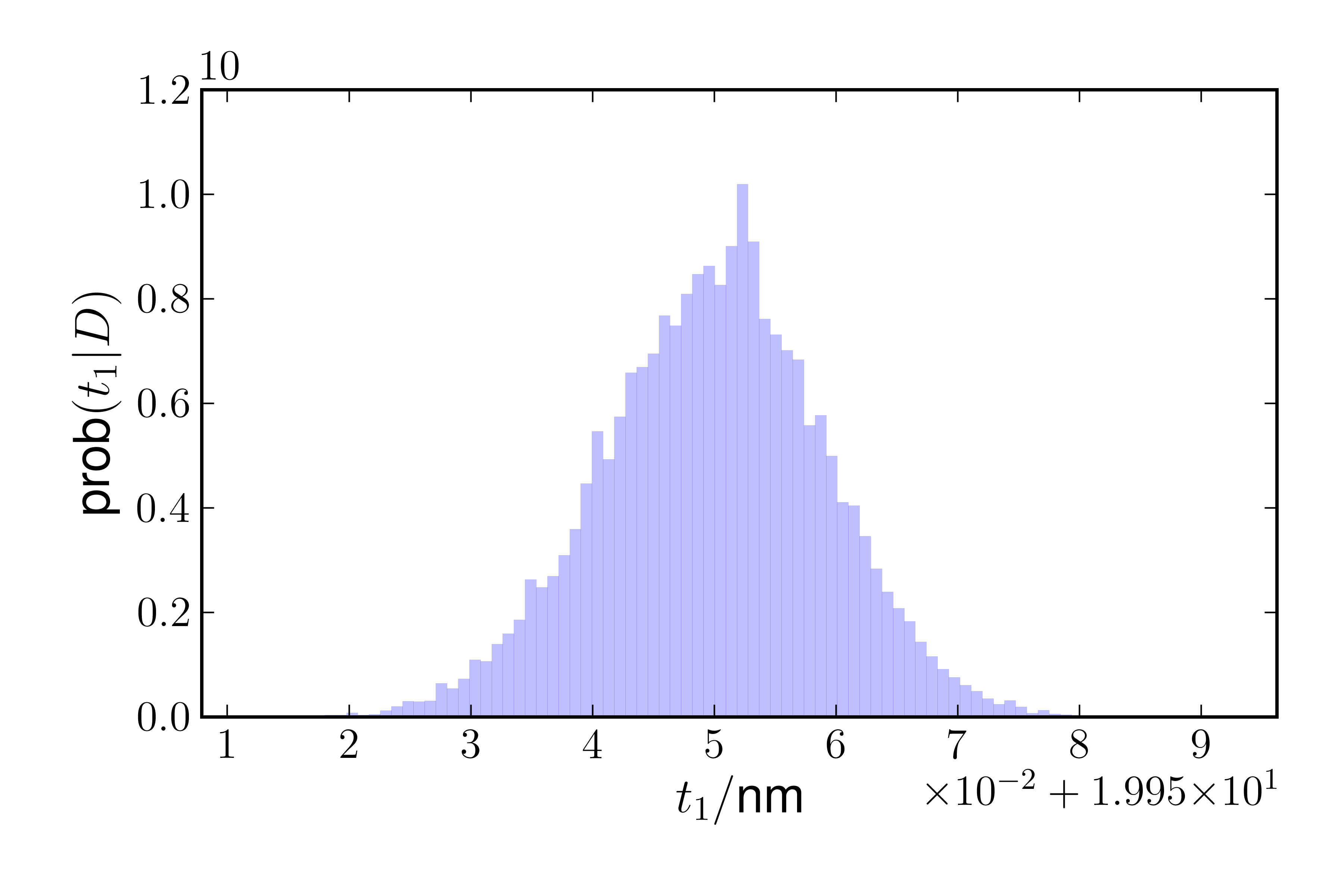}}
  \caption{TiN thickness ({$t_1$}) posterior probability density for case 1 using synchrotron quality data.}
\label{fig:thickness1}
\end{figure}

The power of the Bayesian analysis via MCMC is through its generation of posterior probability distributions for each parameter within a physical model, clearly showing the uncertainty ranges (for example, see Figure \ref{fig:thickness1}).  The expanded uncertainties can be directly calculated by finding the parameter bounds for the probability distribution plot area representing the 95 \%   highest probability.   For case 1, these expanded uncertainty ranges are tabulated in Table \ref{table:case1}.

\begin{figure}
  \resizebox{.45\textwidth}{!}
  	{\includegraphics{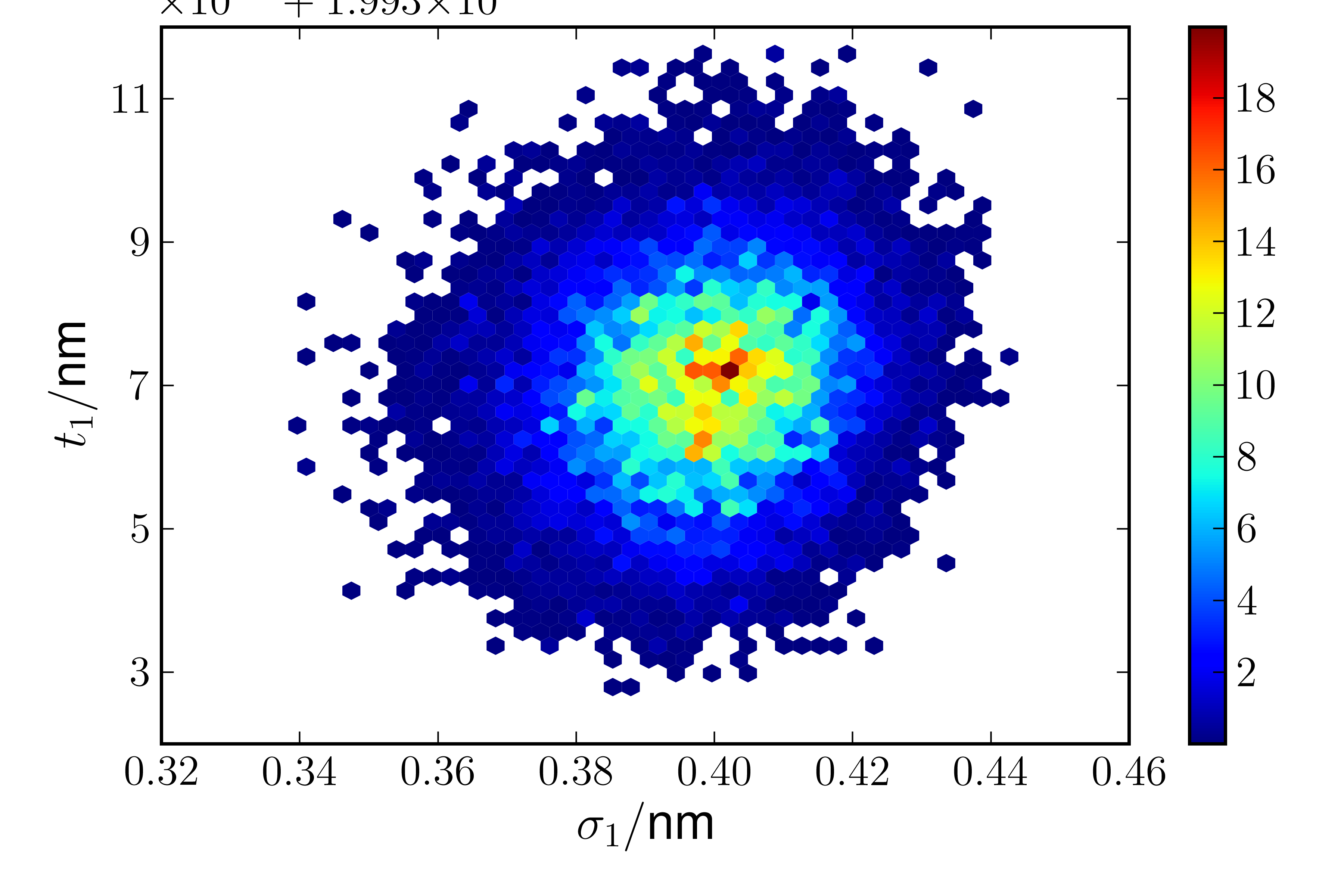}}
  \caption{2-dimensional histogram showing no correlation between TiN thickness ($t_1$) and TiN interface roughness ($\sigma_1$) for case 1.  Intensity scale of histogram shows the relative frequency with which the Monte Carlo Markov Chain explores a given parameter space. (High frequencies directly correspond to high probabilities for a well-tuned MCMC.)}
\label{fig:2dcase1t}
\end{figure}

\begin{figure}
  \resizebox{.45\textwidth}{!}
  	{\includegraphics{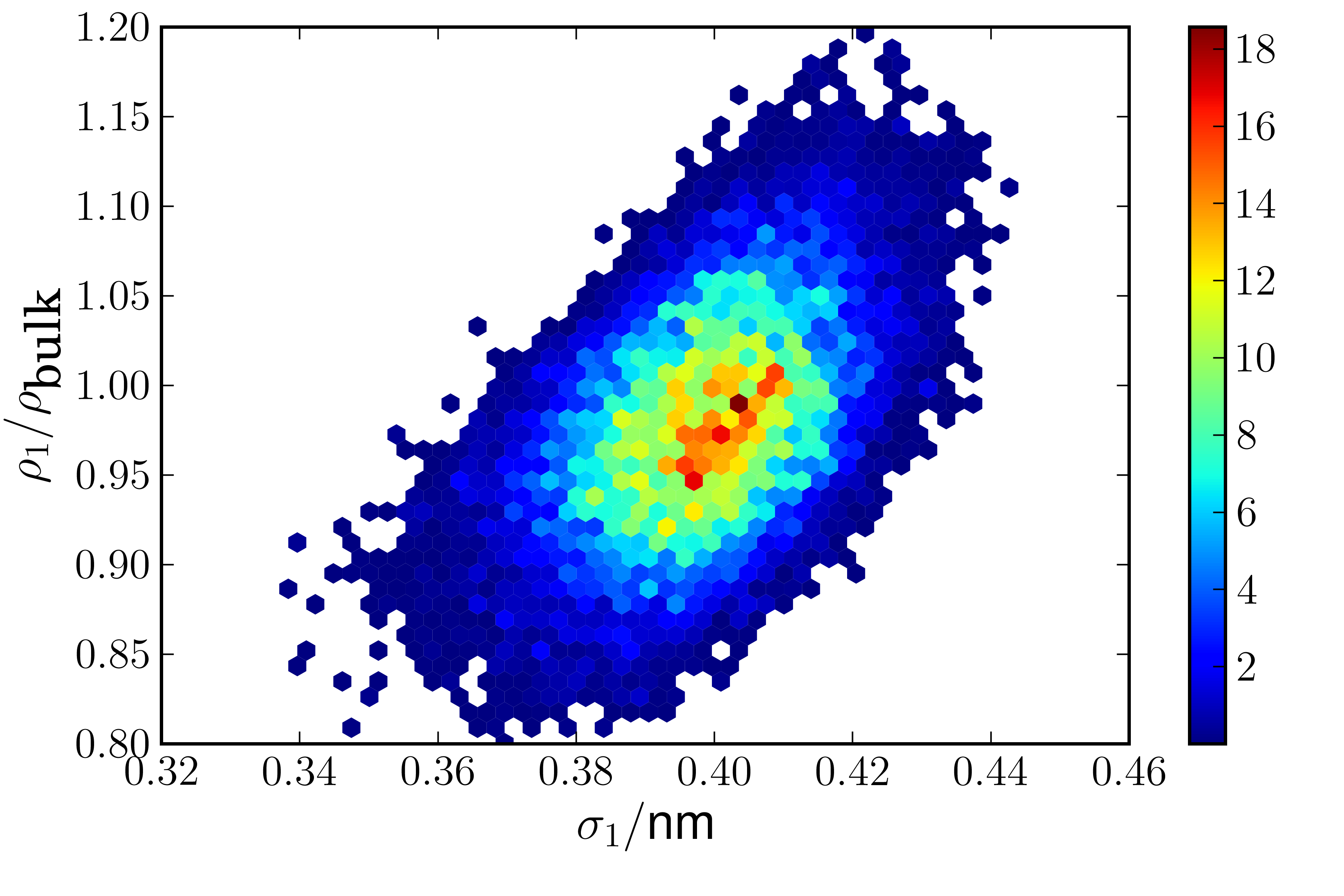}}
  \caption{2-dimensional histogram showing correlation between TiN density ($\rho_1$) and TiN interface roughness ($\sigma_1$) for case 1.   Intensity scale of histogram shows the relative frequency with which the Monte Carlo Markov Chain explores a given parameter space.}
\label{fig:2dcase1r}
\end{figure}

For a clean, single-layer structure, there is a clear (factor of two) advantage to synchrotron measurements with regards to determination of accurate thickness and roughness information.  This statistical determination method for uncertainty estimation is absent from optimization refinement methods such as GAs.  Studying 2-dimensional (2 simultaneous parameters) posterior probability distributions allows us to qualitatively and quantitatively explore parameter correlations.  The clear improvements in TiN thickness and roughness seen in Table \ref{table:case1} are due to the orthogonal (no correlation) nature of thickness and roughness (see Figure \ref{fig:2dcase1t}).  As a general rule, when no correlation exists between parameters within a refinement, then better data quality will directly correspond to reduced uncertainties for the parameters in question.  However, if correlations do exist between two or more parameters, as for example between film roughness and film density (see Figure \ref{fig:2dcase1r}) in case 1, then correlations will introduce intrinsic parameter uncertainties which cannot be further reduced with higher data quality.

As seen by studying the ratios of uncertainty estimates (last column in Table \ref{table:case1}), when determining the density of either the TiN or the Si substrate, there is no clear advantage between the laboratory and the synchrotron levels of data quality.  The relative quality of density determination is nearly identical in both cases (uncertainty ratios equal to 1).  This constant nature of density uncertainty over both levels of data quality is likely caused by two factors: First, that both datasets have the same spacing between collection points, so that the critical angle is not much more precisely determined by synchrotron data than the laboratory data. Second, density correlates with other modeling parameters, for example, interface roughness (see Figure \ref{fig:2dcase1r}).

\begin{table}
\begin{tabular}{cccc}
 \hline
 Parameter
 &U
 &U
 &[U(lab) /
 \\
 &(synchrotron)
 &(laboratory)
 &U(sync)]
 \\
 \hline
$t_\textrm{TiN}$ & \unit{0.048}{\nano\meter}& \unit{0.11}{\nano\meter} & 2.3\\
$\sigma_\textrm{TiN}$ & \unit{0.033}{\nano\meter} & \unit{0.060}{\nano\meter} & 1.8\\
$\sigma_\textrm{Si}$ & \unit{0.050}{\nano\meter} & \unit{0.139}{\nano\meter} & 2.8 \\
$\rho_\textrm{TiN}$ & \unit{1.08}{\gram\per\centi\square\meter} & \unit{1.05}{\gram\per\centi\square\meter} & 1 \\
$\rho_\textrm{Si}$ & \unit{0.90}{\gram\per\centi\square\meter} & \unit{0.90}{\gram\per\centi\square\meter} & 1 \\
\hline
\end{tabular}
\caption{MCMC-determined expanded uncertainties (95\% probability intervals) for the model parameters of case 1, model 1, and the ratio of these uncertainties, $[$U(lab)/U(sync)$]$}
\label{table:case1}
\end{table}

\subsection{Carbon contaminated film - case 2}

\begin{figure}
  \resizebox{.45\textwidth}{!}
  	{\includegraphics{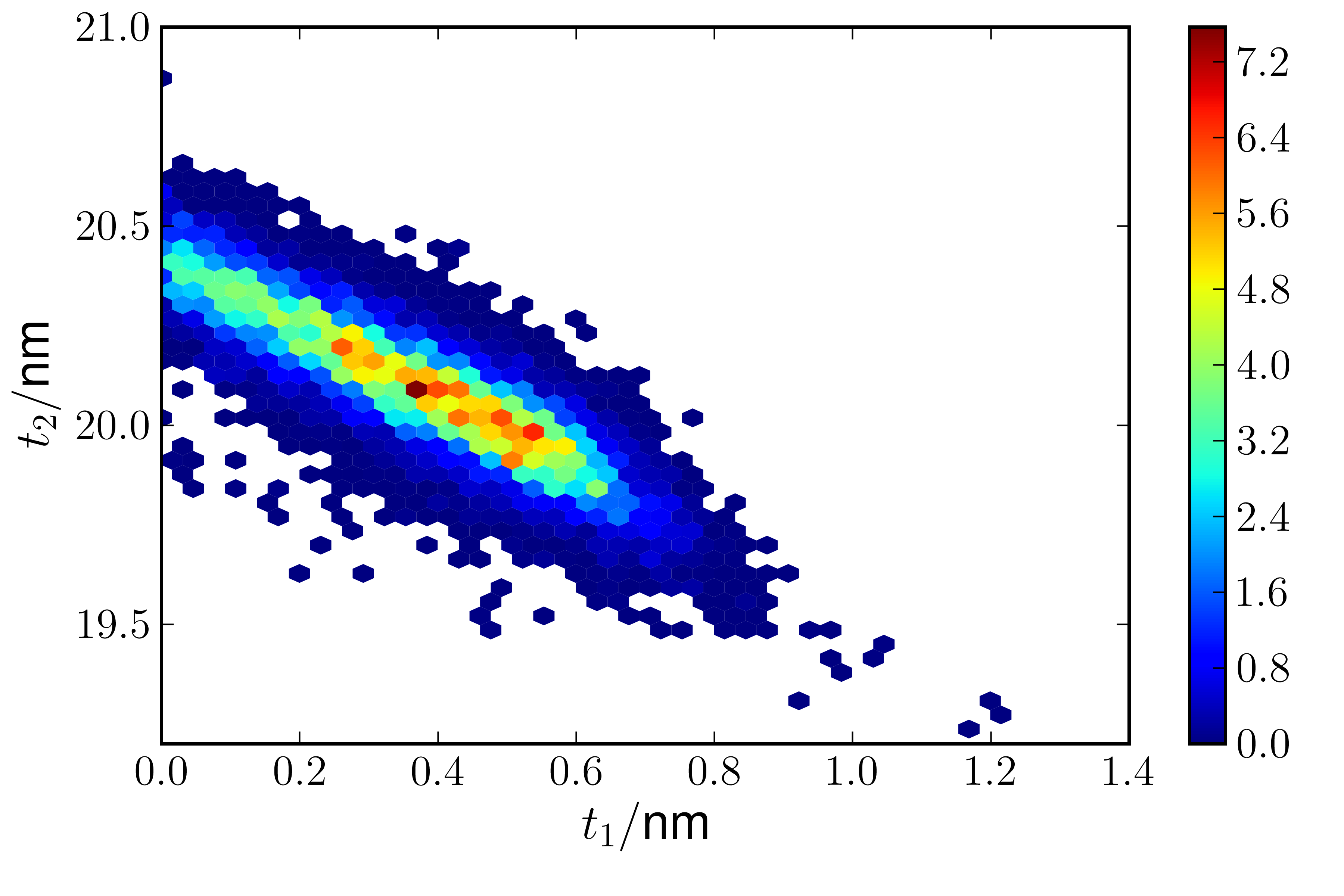}}
  \caption{2-dimensional histogram showing inverse correlation between contamination layer thickness ($t_1$) and TiN thickness ($t_2$) for for case 2.   Intensity scale of histogram shows the relative frequency with which the Monte Carlo Markov Chain explores a given parameter space.}
  \label{fig:2dcase2tt}
\end{figure}

\begin{figure}
  \resizebox{.45\textwidth}{!}
  	{\includegraphics{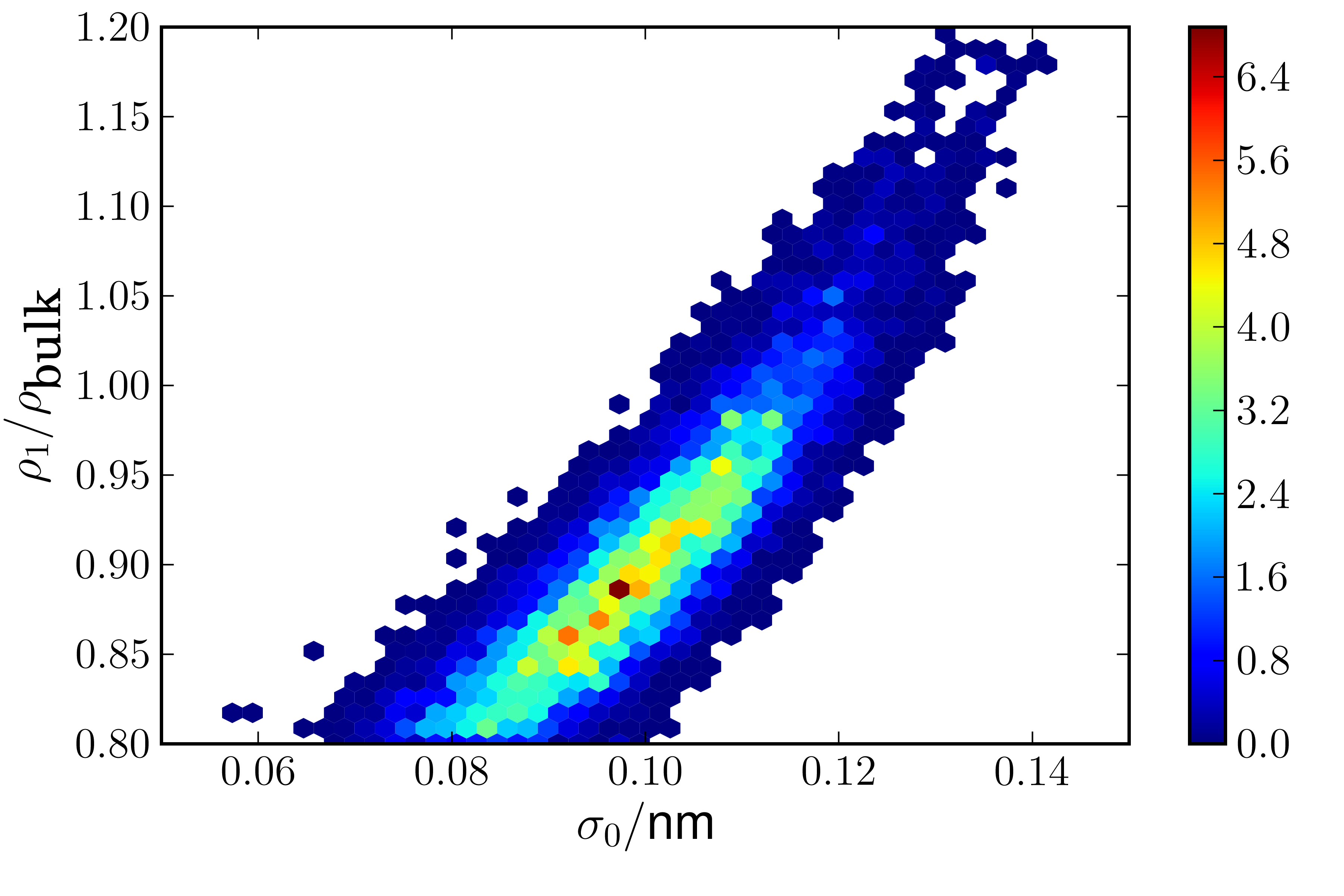}}
  \caption{2-dimensional histogram showing correlation between contamination layer density ($\rho_1$) and surface roughness ($\sigma_0$) for case 2.   Intensity scale of histogram shows the relative frequency with which the Monte Carlo Markov Chain explores a given parameter space.}
  \label{fig:2dcase2dr}
\end{figure}

For case 2, we present the expanded uncertainty ranges in Table \ref{table:case2} and see several surprising results.  When one introduces a carbon contamination layer, the advantages in reduced uncertainties from using the higher data quality of a synchrotron vanishes for all but roughness determination.  For ratios of unity or near unity, (e.g., 0.93, 0.75) there is no clear advantage to synchrotron data. [Apparent disadvantages to synchrotron data (i.e., ratios less than one) are artifacts to the coarseness of our sampling analysis.]  This is partially a consequence of high inverse correlation between contamination layer thickness and TiN thickness (see Figure \ref{fig:2dcase2tt}).  Correlations also exist between the contamination density and surface roughness (see Figure \ref{fig:2dcase2dr}), further expanding uncertainties throughout the model parameters.  

\begin{figure}
  \resizebox{.45\textwidth}{!}
  	{\includegraphics{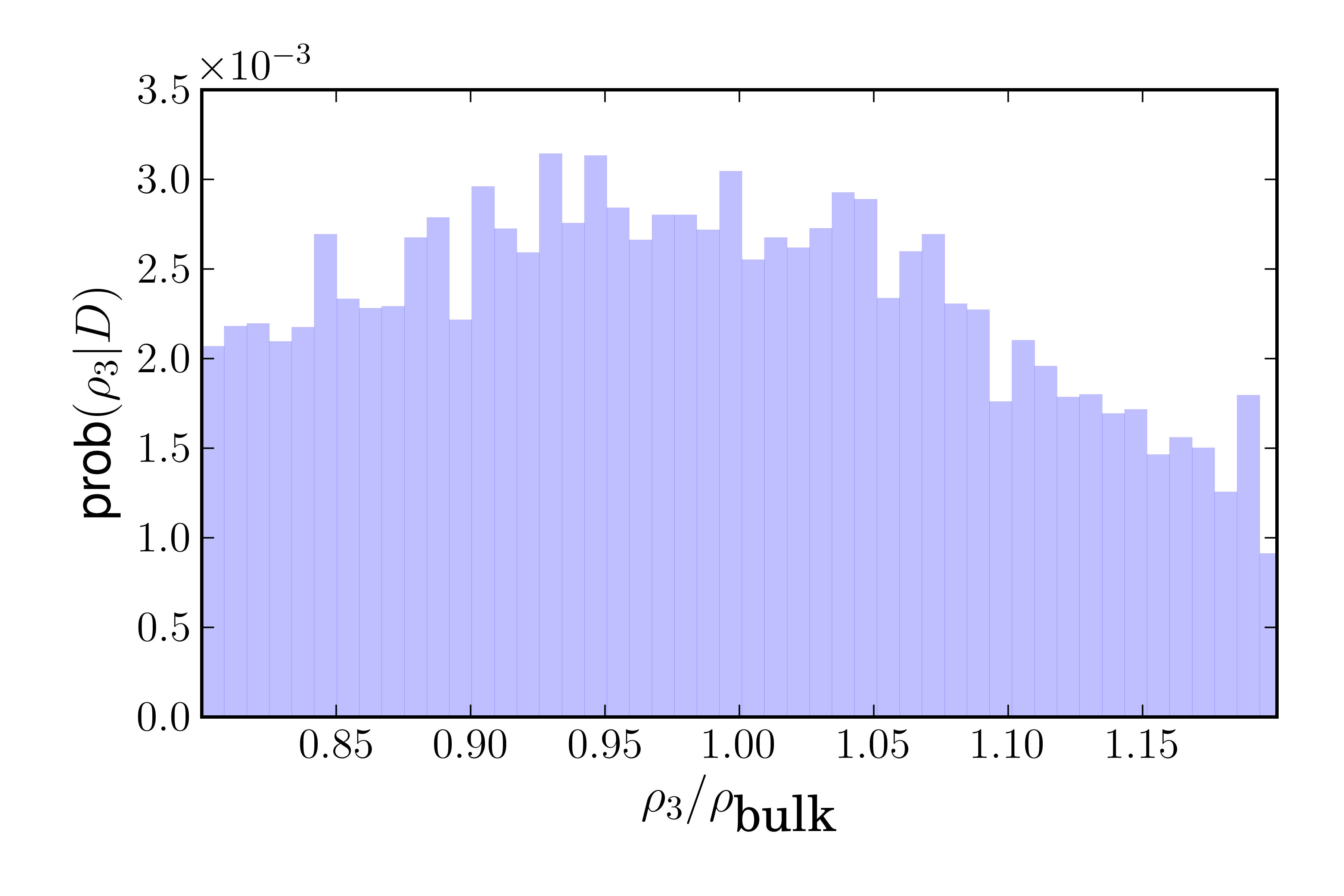}}
  \caption{Posterior probability density for Si substrate density ($\rho_3$) showing lack of sensitivity for this parameter within the XRR model for case 2.}
  \label{fig:2dcase2dt}
\end{figure}

When one examines only the highest quality data (synchrotron), the effect of clean vs. contaminated surfaces can be directly compared.  In Table \ref{table:comparison1},  we see that only the TiN thickness and roughness show pronounced reductions in uncertainty ranges from a contaminated vs. clean structure.  An astute observer may wonder why the uncertainty for Si and TiN density are not improved through the removal of the carbon contamination layer.  This is because XRR, in some cases, is simply not sensitive to a given model parameter.  This sensitivity issue can be distinguished from a correlation phenomena, again by using the MCMC posterior probability densities or 2-dimensional histograms, and looking for parameters which produce uniform posteriors out the analysis.  In Figure \ref{fig:2dcase2dt}, we see that the Si density probability density is nearly uniform over the allowed range of the parameter. This lack of a pronounced peak demonstrates very little sensitivity to Si density in our data.

\subsection{Contamination of known thickness, roughness, and density - case 2a}

In Table \ref{table:comparison2}, we introduce the results from our known parameter carbon contamination case.  We compare the uncertainties between clean, unknown contamination, and exactly known contamination cases   The most interesting feature is the TiN thickness.  Even for the case where the contamination thickness, roughness, and density are known \emph{a priori}, the model still has 5 times higher TiN thickness uncertainty over the clean surface case.  There is a reduction of a factor of 2 over the unknown contamination case; this reduction indicates that not knowing the exact properties of the contamination layer does have an effect on uncertainties. (Or, identically, that refining an unknown contamination layer increases the uncertainties.) But this effect is much smaller than the effect of the mere presence of the contamination layer. This is because the presence of the contamination layer causes a decrease in the contrast of the TiN layer thickness fringes, decreasing the ability of higher quality data to provide more information. This manifests itself in correlations in the model which can substantially increase the overall uncertainties for the TiN layer thickness.

The very presence of contamination on the structure -- rather than the need to fit the contamination -- has the largest effect on the uncertainty.

\section{Conclusions}

There are some caveats to this conclusion: Only simulated data has been considered. All systematic instrumental errors have been neglected. No instrument response functions have been modeled. The comparison between laboratory and synchrotron data is made only for the case of a Cu K$\alpha$ laboratory source radiation and an synchrotron beamline set to the same energy -- so any advantage to tuning the energy of the beam for specific materials and structures has been neglected. (For an example of XRR fit improvement through judicious source energy selection using a synchrotron, see \cite{krumrey_synchrotron_2011}).

But accepting these limitations, save perhaps source energy tuning, as not being likely to \emph{improve} data quality, the MCMC XRR analysis technique provides a powerful tool for studying the theoretical limitations of XRR measurements of a structure before taking measurements.  

In this case of a carbon contamination layer, we see that the theoretical uncertainty estimates for parameters are dominated by correlations between the surface contamination thickness and the TiN thickness increasing the uncertainty estimates for our thin film of interest whenever the carbon contamination layer is present.  Even when armed with prior knowledge of all parameters for the contamination layer, we see a five-fold increase in the TiN thickness uncertainty caused by the introduction of the contamination layer.  Higher data quality will provide significant reductions in parameter uncertainties for simple XRR models, such as the clean TiN thin film.  However, in the presence of contamination, we see minimal gain through enhanced data quality for the determination of thin film thickness.

This MCMC simulated data study has shown that removing the contamination is essential to significantly reducing the uncertainties in the high-Z layer thickness measurement.

\begin{table}
\begin{tabular}{lccc}
 \hline
 Parameter
 &U
 &U
 &[U(lab) /
 \\
 &(synchrotron)
 &(laboratory)
 &U(sync)]
 \\
 \hline
$t_\textrm{C}$ & \unit{0.64}{\nano\meter}&  \unit{0.66}{\nano\meter} &  1.0\\
$t_\textrm{TiN}$ & \unit{0.67}{\nano\meter} & \unit{0.62}{\nano\meter} & 0.93\\
$t_\textrm{C} + t_\textrm{TiN}$ & \unit{0.32}{\nano\meter} & \unit{0.24}{\nano\meter} & 0.75\\

$\sigma_\textrm{C}$ & \unit{0.033}{\nano\meter} & \unit{0.16}{\nano\meter} & 4.8\\
$\sigma_\textrm{TiN}$ & \unit{0.50}{\nano\meter} & \unit{0.48}{\nano\meter} & 0.96\\
$\sigma_\textrm{Si}$ & \unit{0.027}{\nano\meter} & \unit{0.20}{\nano\meter} & 7.4\\

$\rho_\textrm{C}$ & \unit{0.61}{\gram\per\centi\square\meter}& \unit{0.92}{\gram\per\centi\square\meter} & 1.5 \\
$\rho_\textrm{TiN}$ & \unit{1.49}{\gram\per\centi\square\meter} & \unit{1.28}{\gram\per\centi\square\meter} & 0.86 \\
$\rho_\textrm{Si}$ & \unit{0.94}{\gram\per\centi\square\meter} & \unit{0.94}{\gram\per\centi\square\meter} & 1 \\

\hline
\end{tabular}
\caption{MCMC-determined expanded uncertainties (95\% probability intervals) for the model parameters of case 2, model 2, and the ratio of these uncertainties, $[$U(lab)/U(sync)$]$.}
\label{table:case2}
\end{table}

\begin{table}
\begin{tabular}{lccc}
 \hline
 Parameter
 &U(clean)
 &U(with carbon)
 &[U(carbon) /
 \\
 &(synchrotron)
 &(synchrotron)
 &U(clean)]
 \\
 \hline
$t_\textrm{TiN}$ & \unit{0.048}{\nano\meter}& \unit{0.67}{\nano\meter} & 14\\
$\sigma_\textrm{surface}$ & \unit{0.033}{\nano\meter} & \unit{0.033}{\nano\meter} & 1\\
$\sigma_\textrm{TiN}$ & \unit{0.033}{\nano\meter} & \unit{0.5}{\nano\meter} & 15\\
$\sigma_\textrm{Si}$ & \unit{0.050}{\nano\meter} & \unit{0.027}{\nano\meter} & 0.54 \\
$\rho_\textrm{TiN}$ & \unit{1.49}{\gram\per\centi\square\meter} & \unit{1.49}{\gram\per\centi\square\meter} & 1 \\
$\rho_\textrm{Si}$ & \unit{0.94}{\gram\per\centi\square\meter} & \unit{0.94}{\gram\per\centi\square\meter} & 1 \\
\hline
\end{tabular}
\caption{Comparison of MCMC-determined expanded uncertainties (95\% probability intervals) between clean vs. contaminated cases for synchrotron quality data}
\label{table:comparison1}
\end{table}

\begin{table}
\begin{tabular}{lccc}
 \hline
 Parameter
 &U(clean)
 &U(carbon)
 &U(known
 \\
 &
 &
 &carbon)
 \\
 \hline
$t_\textrm{TiN}$ & \unit{0.048}{\nano\meter}& 
\unit{0.67}{\nano\meter} & 
\unit{0.31}{\nano\meter}\\
$\sigma_\textrm{surface}$ & \unit{0.033}{\nano\meter} & 
\unit{0.033}{\nano\meter} & 
fixed\\
$\sigma_\textrm{TiN}$ & same as $\sigma_\textrm{surface}$& 
\unit{0.5}{\nano\meter} &
\unit{0.36}{\nano\meter}  \\
$\sigma_\textrm{Si}$ & \unit{0.050}{\nano\meter} & 
\unit{0.027}{\nano\meter} & 
\unit{0.027}{\nano\meter} \\
$\rho_\textrm{TiN}$ & \unit{1.49}{\gram\per\centi\square\meter} & 
\unit{1.49}{\gram\per\centi\square\meter} & 
\unit{1.49}{\gram\per\centi\square\meter} \\
$\rho_\textrm{Si}$ & 
\unit{0.94}{\gram\per\centi\square\meter} & 
\unit{0.94}{\gram\per\centi\square\meter} & 
\unit{0.93}{\gram\per\centi\square\meter} \\
\hline
\end{tabular}
\caption{Comparison of MCMC-determined expanded uncertainties (95\% probability intervals) between clean, unknown, and known contamination layer cases for synchrotron quality data}
\label{table:comparison2}
\end{table}


\ack
We would like to thank Victor Vartanian of the International SEMATECH Manufacturing Initiative (ISMI) (Albany, NY) for providing pre-standard test structures and XRR measurements for XRR SRM development at NIST.  We would also like to thank P.Y. Hung of Sematech (Albany, NY) for extensive and very helpful discussions on thin film characterization and for providing numerous interesting sets of XRR data for the development of analysis techniques.  

\section*{References}

\bibliographystyle{unsrt}   

\bibliography{07_MRS_windover_mod,2009_frontiers_xrr}

 {\typeout{}
  \typeout{******************************************}
  \typeout{** Please run "bibtex \jobname" to obtain}
  \typeout{** the bibliography and then re-run LaTeX}
  \typeout{** twice to fix the references!}
  \typeout{******************************************}
  \typeout{}
 }

\end{document}